\documentclass[fleqn,usenatbib]{mnras}
\usepackage[T1]{fontenc}

\DeclareRobustCommand{\VAN}[3]{#2}
\let\VANthebibliography\thebibliography
\def\thebibliography{\DeclareRobustCommand{\VAN}[3]{##3}\VANthebibliography}

\usepackage{graphicx}	
\usepackage{amsmath}	
\usepackage{amssymb}	
\usepackage{bm}         
\usepackage{tikz}

\newcommand{\lya}{\mbox{Lyman-$\alpha$}}

\newcommand{\at}{\textsc{aton}}

\newcommand{\atmf}{\textsc{aton-he}}

\definecolor{notecolor}{rgb}{0.8,0,0}

\usepackage{newtxtext,newtxmath}

\newcommand{\xion}{$\xi_\text{ion}$}
\newcommand{\fesc}{$f_\text{esc}$}

\title[Photon budget during reionization]{The ionizing photon budget and effective clumping factor in radiative transfer simulations calibrated to \lya\ forest data}
\author[Asthana et al.]{
Shikhar Asthana$^{1}$\thanks{E-mail: sa2001@cam.ac.uk},
Girish Kulkarni$^{2}$,
Martin G. Haehnelt$^{1}$, 
James S. Bolton$^{3}$ \newauthor
Laura C. Keating$^{4}$,
and Charlotte Simmonds$^{1,5}$ 
\\
$^{1}$Kavli Institute for Cosmology and Institute of Astronomy, Madingley Road, Cambridge, CB3 0HA, UK\\
$^{2}$Tata Institute of Fundamental Research, Homi Bhabha Road, Mumbai 400005, India\\
$^{3}$School of Physics and Astronomy, University of Nottingham, University Park, Nottingham, NG7 2RD, UK\\
$^{4}$Institute for Astronomy, University of Edinburgh, Blackford Hill, Edinburgh, EH9 3HJ, UK\\
$^{5}$Cavendish Laboratory, University of Cambridge, 19 JJ Thomson Avenue, Cambridge, CB3 0HE, UK\\
}
\date{Accepted ---. Received ---; in original form ---}
\pubyear{2023}
\begin{document}
\label{firstpage}
\pagerange{\pageref{firstpage}--\pageref{lastpage}}
\maketitle
\begin{abstract}
Recent JWST observations have allowed for the first time to obtain comprehensive measurements of the ionizing photon production efficiency $\xi_\text{ion} $ for a wide range of reionization-epoch galaxies. We explore implications for the inferred UV luminosity functions and escape fractions of ionizing sources in our suite of simulations. These are run with the GPU-based radiative transfer code \atmf\ and are calibrated to the XQR-30 \lya\ forest data at $5<z<6.2$. For our fiducial source model, the inferred ionizing escape fractions increase from (6.1, 5.4, 4.9)\% at $z=6$ to (14.4, 23.8, 29.4)\% at $z=10$ for our (Fiducial, Early, Extremely Early) models in good agreement with extrapolations of lower redshift escape fraction measurements. Extrapolating observed luminosity functions beyond the resolution limit of the simulations to faint sources with $M_\text{UV}=-11$ increases the inferred escape fractions by a factor $\sim 1.5$ at $z=10$. For our oligarchic source model, where no ionizing photons are emitted in faint sources, the inferred escape fractions increase from 10\% at $z=6$ to uncomfortably large values $>50$\% at $z> 10$,  disfavouring the oligarchic source model at very high redshift. The inferred effective clumping factors in our simulations are in the range of $3-6$, suggesting consistency between the observed ionizing properties of reionization-epoch galaxies and the ionizing photon budget in our simulations.  
\end{abstract}
\section{Introduction}
The Epoch of Reionization (EoR) is a transformative phase in cosmic history when the intergalactic medium (IGM) transitioned from neutral to ionized due to the emergence of the first luminous sources \citep{Mcquinn2016, Dayal2018}. Several observational probes, including the cosmic microwave background (CMB) Thomson scattering optical depth \citep{Kogut2003, Planck2020}, and \lya\ forest observations, have been pivotal in constraining the timeline of reionization \citep{Fan2006, Mcgreer2014, Kulkarni2019, Keating2019,  Bosman2018, Bosman2022, Becker2021, Dodorico2023}, indicating that reionization ends at $z\lesssim5.5$. The sources that cause reionization, however, are still uncertain, with ongoing discussions focusing on low-mass faint galaxies, bright, massive galaxies \citep{Finkelstein2019, Naidu2020, Yeh2023}, and active galactic nuclei \citep[AGN;][]{Dayal2024, Madau2024, Asthana2024-2}.

\begin{figure*}
    \includegraphics[width=\linewidth]{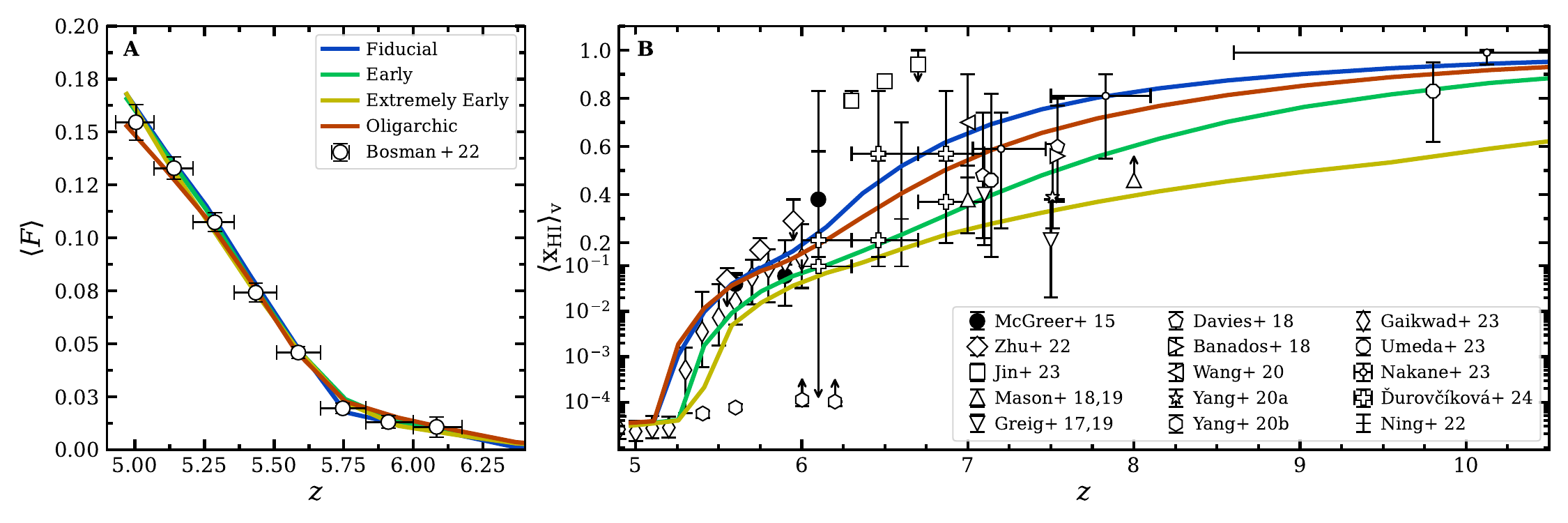}
    \caption{Panel~A compares the mean \lya\ forest transmission, $\langle F\rangle$, in our simulations, with measurements by \citet{Bosman2022}.  Panel~B shows the volume-averaged neutral hydrogen fraction, $\langle x_\mathrm{HI}\rangle_\mathrm{v}$.  This panel also shows inferences of the neutral hydrogen fraction from various observations: the fraction of Lyman-break galaxies showing \lya\ emission \citep{Mason2018, Mason2019}, dark gaps in the \lya\ forest \citep{Mcgreer2014, Zhu2022, Jin2023}, \lya\ emission equivalent widths \citep{Nakane2023}, quasar damping wings \citep{Greig2017, Banados2018, Davies2018, Greig2019, Wang2020, Yang2020, Durovcikova2024}, the effective \lya\ opacity of the IGM \citep{Yang2020-2, Ning2022, Gaikwad2023}, and galaxy damping wings \citep{Umeda2023}.}
    \label{fig:reionizaiton_history}
\end{figure*}

Recent observations by the James Webb Space Telescope (JWST) have allowed one to measure the ionizing photon production efficiency ($\xi_\text{ion}$) for a wide range of reionization-epoch galaxies for the first time. The first studies suggested an unexpectedly large evolution of \xion with redshift \citep{Atek2024, Simmonds2024a}, a much steeper slope than previous estimates. For larger samples that better account for selection effects, the evolution is more moderate \citep{Simmonds2024b, Begley2024}. Combining the first set of values with estimates of the escape fraction, \citet{Munoz2024} suggested that there may be too many ionizing photons for reionization to end as late as indicated by \lya\ forest data. This argument by \citet{Munoz2024} also involved extrapolating the UV luminosity density to fainter luminosities and high redshifts and assumed an evolution of the effective clumping factor of the IGM based on numerical simulations \citep{Chen2020}. This discrepancy became known as the ``photon budget crisis'' \citep{Munoz2024}. To resolve this tension, \citet{Davies2024} proposed that a proportionately greater number of recombinations offsets the increased number of ionizing photons suggested by JWST observations. As a result, they inferred an effective clumping factor as high as 15 during the late stages of reionization. 

This motivates us to closely examine the photon budget in our reionization models, using GPU-based cosmological radiative transfer simulations carefully calibrated to \lya\ forest data \citep{Asthana2024, Asthana2024-2}.

The letter is structured as follows. Section \ref{sec:simset} outlines our simulation set-up. Section \ref{sec:history} presents results from our models, focusing on the reionization history, UV luminosity function, ionization photon production efficiency (\xion), escape fraction (\fesc), and effective clumping factor. Section \ref{sec:conclusion} summarizes our findings and concludes the letter. We adopt a $\Lambda$CDM cosmology with parameter values $\Omega_\mathrm{m} = 0.308$, $\Omega_\lambda = 0.6982$, $h = 0.678$, $\Omega_\mathrm{b} = 0.0482$, $\sigma_8 = 0.829$, and $n_\mathrm{s} = 0.961$ \citep{Planck2014}.

\begin{figure*}
    \centering
    \includegraphics[page=1,width=\linewidth]{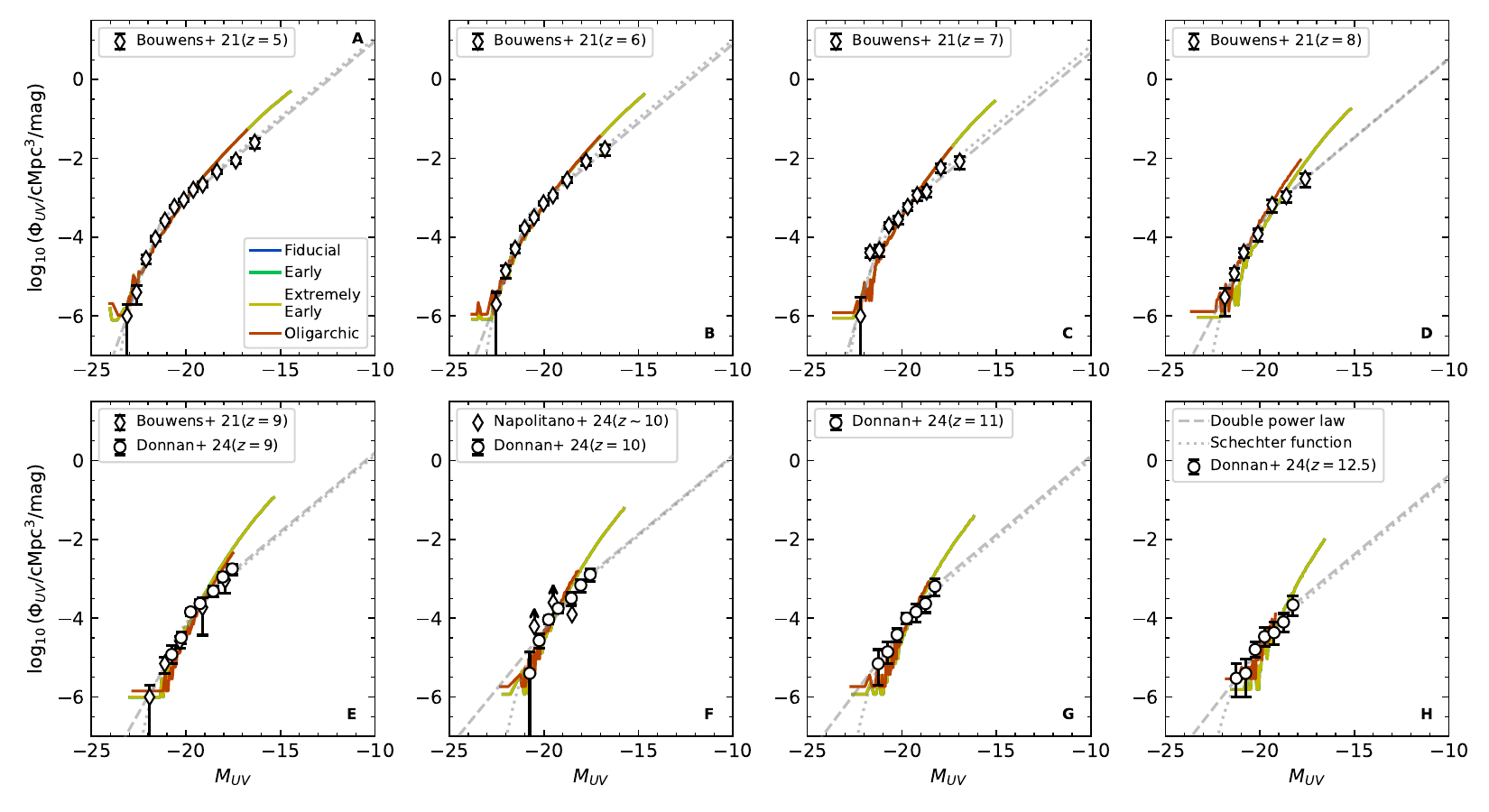}
    \caption{The UV luminosity function in our four models at redshifts  $z\,=\,5.11,\,5.95,\,7.14,\,8.15,\,9.02,\,10.14,\,10.83,\,12.59$. The observational data points are taken from \citet{Bouwens2021}, \citet{Donnan2024}, and \citet{Napolitano2024}. The dashed and dotted grey curves represent the best-fit Schechter function and double power law (with faint end slope $\propto L^{-1}$) curves, respectively, to the data points. 
    }
    \label{fig:UV_luminosity}
\end{figure*}

\section{Simulation set-up}\label{sec:simset}
Our simulation set-up is described in detail in \citet{Asthana2024, Asthana2024-2}. We summarize the essential details here.

The simulations are carried out using the GPU-based M1-closure multi-frequency radiative transfer code \atmf\ \citep{Asthana2024}, which is a modified version of the \at\ code \citep{Aubert2008, Aubert2010}, by post-processing cosmological hydrodynamical simulations from the Sherwood-Relics suite of simulations \citep{Puchwein2023}.  \atmf\ tracks the ionization states of both hydrogen and helium. The Sherwood-Relics simulations were performed using the Tree-PM SPH code \mbox{\textsc{p-gadget-3}}. The simulations include $2\times2048^3$ gas and dark matter particles in a 160~cMpc$/h$ box. The simulations start at $z=99$, with snapshots saved every $40$~Myr down to $z=4$. Star formation is modelled using a simplified prescription. enabled by the \texttt{QUICK\_LYALPHA} compile-time flag in \mbox{\textsc{p-gadget-3}}, where gas particles that exceed a density threshold of $\Delta = 10^3$, and have temperature $< 10^5$~K, are converted into star particles \citep{Viel2004}.  A uniform UV background as described by \citet{Puchwein2019} is integrated into the simulations to approximate the hydrodynamic response of the gas density to reionization.

For the post-processing, gas density is projected onto a uniform Cartesian grid of $2048^3$ cells. 
Once the matter distribution is established, ionizing sources are placed at the locations of dark matter haloes following a source model. The total ionizing emissivity in the simulation is treated as a free parameter, adjusted to fit the mean \lya\ forest transmission at redshifts $5\lesssim z \lesssim 6.2$ \citep{Bosman2022}. To do this, we calculated the mean transmission along 6400 sightlines using the Voigt profile approximation by \citet{Thorsten2006}. Based on this, in \citet{Asthana2024} we have explored three models with different reionization mid-points: the fiducial ($z_\mathrm{mid}\sim6.5$), `Early' ($z_\mathrm{mid}\sim7.5$), and `Extremely Early' models ($z_\mathrm{mid}\sim8.5$) where $z_\mathrm{mid}$ is the redshift at the midpoint of reionization. Furthermore, by setting the minimum halo mass that emits ionizing photons to $8.5\times10^9~\text{M}_\odot/h$ (as opposed to $10^9~\text{M}_\odot/h$ for the other models), we further explored an  `Oligarchic' model similar to that in \citet{Cain2023}. We discuss the reionization histories of these four models in the next section.

\section{Results}\label{sec:history}
\subsection{Reionization histories}

In Panel~A of Figure~\ref{fig:reionizaiton_history}, we show the mean \lya\ transmission in our simulations compared to the data, indicating our calibration. The `Fiducial', `Early', `Extremely Early' and `Oligarchic' models are shown in this figure by the blue, green, yellow and red curves, respectively. The simulations match the observational measurements by \citet{Bosman2022} very well. In Panel~B of Figure~\ref{fig:reionizaiton_history}, we show the reionization history of the four models, together with a large number of inferences from the literature. As suggested by the name, the `Extremely Early' model has the highest redshift for the midpoint for reionization, while the fiducial model has the lowest.

\subsection{UV luminosity functions}

As discussed above, the volume ionizing emissivity was chosen to match the observed mean Lyman-$\alpha$ transmission. The volume emissivity is distributed over the dark matter haloes identified in the simulation in proportion to the halo mass, as described in \citet{Asthana2024}. As in \citet{Asthana2024-2}, we can then map this ionizing emissivity $\dot N_\text{ion}$ of each source to a UV luminosity $L_{\text{UV}}$ by assuming values for the product of ionizing photon production efficiency $\xi_\text{ion}$ and the LyC escape fraction $f_\text{esc}$ for ionizing photos, as
\begin{equation}
    L_{\text{UV}} = \frac{\dot N_\text{ion}}{f_\text{esc}\xi_\text{ion}}.
    \label{eq:L_UV}
\end{equation}
This allows us to compare the UV luminosity function of galaxies and AGN in our models with observations.

We calculate the absolute UV magnitude $M_\text{UV}$ at 1450~\AA\ using the relation \citep{1983ApJ...266..713O},
\begin{equation}
  M_{\text{UV}} = -2.5\log_{10}\left(\frac{L_{\text{UV}}}{\text{erg s}^{-1}\text{Hz}^{-1}}\right)+51.63,
    \label{eq:UV_luminosity}
\end{equation}
where $L_\text{UV}$ is the UV luminosity at the same wavelength. 
Then, using Equation~(\ref{eq:L_UV}), we can write,
\begin{multline}
  M_{\text{UV}} = -19.62 -2.5\log_{10}\left(\frac{\dot N_\text{ion}}{10^{53}\text{s}^{-1}}\right)\\+2.5\log_{10}\left(\frac{f_\text{esc}}{0.1}\right)+2.5\left(\log_{10}\xi_\text{ion}-25.5\right).
  \label{eq:full_muv_ndot}
\end{multline}
The resulting luminosity functions for our four models are shown in Figure~\ref{fig:UV_luminosity} for redshifts $z=5$--$13$, along with observational measurements from \citet{Bouwens2021}, \citet{Donnan2024}, and \citet{Napolitano2024}.  The dashed and dotted grey lines represent the best-fit double-power-law and Schechter functions (with faint end slope $\propto L^{-1}$), respectively, when fit to the observational data points. To achieve agreement with the observed luminosity functions, we have adjusted the combined value of $f_\text{esc}\times\xi_\text{ion}$ and obtained a least square fit independently at each redshift.  For the fit, we keep $f_\text{esc}\times\xi_\text{ion}$ fixed, independent of the halo mass of the ionizing sources.   

The UV luminosity function inferred for the simulations agrees with the observations at the bright end.  The faint end is somewhat steeper than a $L^{-1}$ extrapolation of the observed data, especially at high redshift. Note here that in the simulations, we do not have a physical model for the suppression of star formation in small mass haloes. The halo mass cutoff at $10^9\,\text{M}_\odot/h$ for the Fiducial, Early, and Extremely Early models is set by the mass resolution of the post-processed Gadget simulation. The halo mass cutoff of $8.5\times10^9\,\text{M}_\odot/h$ in the oligarchic model is manually imposed, causing the red line to terminate earlier than in the other models. 

\begin{figure*}
    \centering
    \includegraphics[page=2, width=\linewidth]{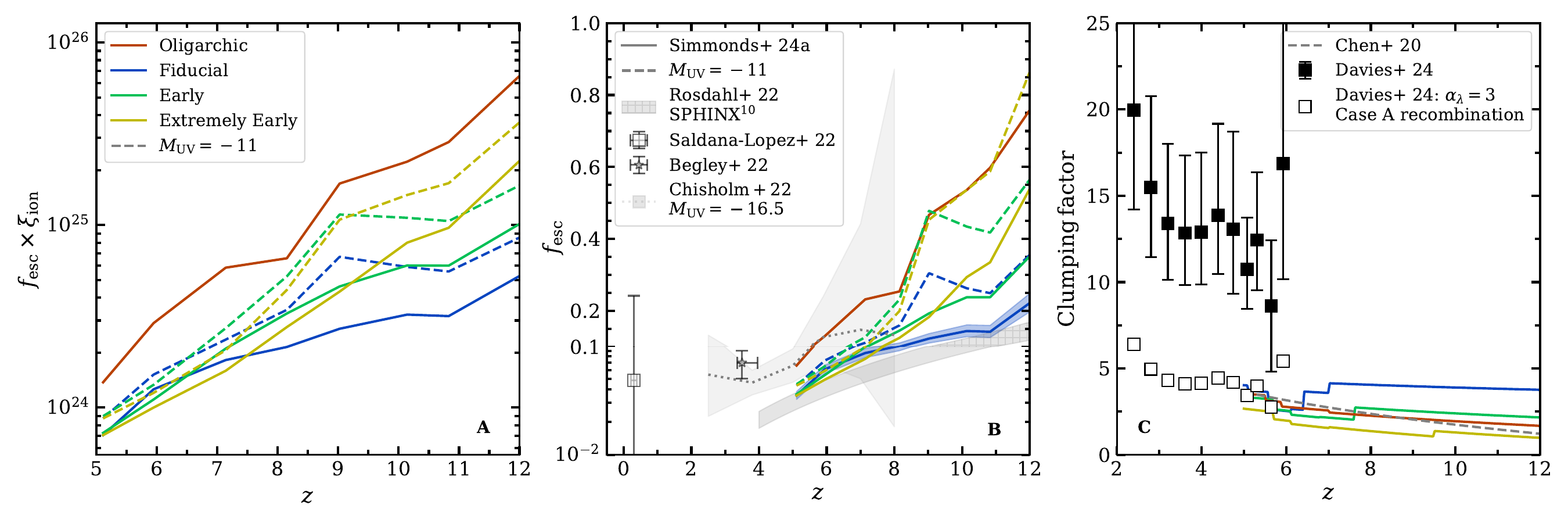}
    \caption{In Panel~A, the solid lines represent the values of $f_\text{esc}\times\xi_\text{ion}$ required to match the observed UV luminosity function for the four models. The dashed line represents the same value now integrating the observed UV LF down to $M_\text{UV}=-11$. For the oligarchic model, an extension to fainter magnitudes is not sensible as this model is constructed to model reionization by bright sources only. In Panel~B, the solid lines represent the evolution of the escape fraction with redshift, assuming a value of $\xi_\text{ion}$ from \citet{Simmonds2024b}. In contrast, the dashed lines show the escape fraction if the observed UV LFs are integrated to $M_\text{UV}=-11$. The star shows the constraint from \citet{Begley2022} while the dark grey band indicates the range of values found by the two SPHINX$^{10}$ simulations \citep{Rosdahl2022}. The square with the error bar shows the mean value and the range for the escape fraction reported by \citet{2022A&A...663A..59S}. The light grey curve is the constraints on the escape fraction found by fitting a Schechter function through the observed UV LFs and using the relation between escape fraction and $M_\text{UV}$ given in \citet{Chisholm2022}, with normalization constants taken from in \citet{Bouwens2014}. In Panel~C, the solid lines show the clumping factor from our models, estimated as described in Section~\ref{sec:clump}. The dashed grey curve is the evolution of the clumping factors from \citet{Chen2020}, while the solid points are the estimates from \citet{Davies2024}. The open squares are the values from \citet{Davies2024}, modified by replacing the case~B recombination rate used in their analysis with its case~A counterpart, and divided by a further factor 1.88 assuming $\alpha_\lambda=3$ instead of 1.}
    \label{fig:Clumping}
\end{figure*}

\subsection{Ionizing production efficiency and escape fraction}

As discussed in the last section, to match the observed luminosity function, we modulate the factor $f_\text{esc}\times \xi_\text{ion}$. The value of this factor used to match the luminosity function in Figure~\ref{fig:UV_luminosity} is shown by solid curves in panel~A of Figure~\ref{fig:Clumping}. Note that we assume $f_\text{esc}\times\xi_\text{ion}$ to be independent of the mass of the host haloes of the ionizing sources. 

The UV luminosity functions inferred from our simulations only extend to moderately faint sources with $M_\text{UV} = -14.5$ at $z\sim5$ for our fiducial source model. To explore the possible effect of faint objects not resolved in our models, we therefore also calculate the inferred evolution of $f_\text{esc}\times\xi_\text{ion}$ extending the observed UV luminosity function to $M_\text{UV} = -11$ \citet{Munoz2024}
using the best-fit Schechter function as shown by the dotted line in Figure~\ref{fig:UV_luminosity}. 
As the faint end of the inferred luminosity function in our simulation is steeper than $L^{-1}$, this increases the inferred  $f_\text{esc}\times\xi_\text{ion}$ by a factor of 1.15 despite the extrapolation to fainter magnitudes.  This is shown as the dashed curves in panel~A of Figure~\ref{fig:Clumping}. Note that these values are also similar to the ones found in the Lyman-$\alpha$ calibrated ray-tracing RT simulations by \citet{Cain2024}. For the oligarchic model, an extension to fainter magnitudes is not sensible as this model is constructed to model reionization by bright sources only.  

As discussed in the introduction, measurements of the ionizing photon production efficiency for a wide 
range of reionization-epoch galaxies are now available. This allows us to turn the $f_\text{esc}\times\xi_\text{ion}$ evolution into a corresponding evolution of $f_\text{esc}$ shown in Panel~B of Figure~\ref{fig:Clumping}. For this, we use the photometric measurements of galaxies between $3<z<9$ \citep{Simmonds2024b},  neglecting any possible dependence on luminosity. This assumption is motivated by the results of \citet{Begley2024}, who found a weak dependence of \xion on magnitude. Note that the measurements 
 of  \citet{Simmonds2024b} are averaged over galaxies with a wide range of star-forming properties and have a relatively flat evolution with redshift. 

As we see in panel~B of Figure~\ref{fig:Clumping}, the inferred average escape fraction decreases rapidly with decreasing redshift and becomes uncomfortably high for our oligarchic source model at high redshift. This is also true for the Extremely Early model, albeit at a somewhat higher redshift. The blue band represents the uncertainty of the inferred escape fractions taking the upper and lower limits of the \xion\ relation in \citet{Simmonds2024b}. The uncertainty is similar for all the simulations, so we show it only for the fiducial model. In the cases where we use the UV luminosity functions extended to faint magnitudes, we see that the escape fraction increases by about a factor of 1.5 at higher redshifts and is fairly similar towards the end of reionization. At redshifts $z\sim 5$--$6$, the inferred escape fraction falls below 10\%, in good agreement with direct measurements at lower redshift, perhaps even on the low side. \citet{Begley2022}, e.g., obtained a value of $7\pm 2 $ \% at $z\sim 3.5$.  Note further that for the QSO-assisted models in \citet{Asthana2024-2}, the ionizing volume emissivity at $z<10$ is up to a factor 1.8 lower than in our fiducial model, and the inferred escape fractions would be correspondingly lower. In light grey, we show the escape fractions inferred from the observed relationship between escape fraction and magnitude at low redshift   \citep{Chisholm2022}. 

We also show the evolution of the escape fraction for the two SPHINX$^{10}$ \citep{Rosdahl2022} simulations, with a box size of 10~Mpc$/h$, that straddle the late stages of the reionization history suggested by \lya\ forest data.\footnote{The SPHINX$^{20}$  simulation with a box size of 20 Mpc$/h$  completes reionization somewhat later than suggested by the \lya\ forest data and has somewhat lower escape fractions.}  It shows a similar decrease with decreasing redshift as our simulations with the fiducial source model. Note that the SPHINX simulations have significantly higher resolution and smaller box sizes, and the luminosity evolution of the ionizing sources is rather bursty.  

\subsection{Effective clumping factor}
\label{sec:clump}

\citet{Madau1999} modelled the evolution of the volume-filling factor of ionized hydrogen, $Q_\text{HII}$, as
\begin{equation}
    \frac{dQ_\text{HII}}{dt} = \frac{\dot{n}_\text{ion,H}}{\langle n_\text{H} \rangle} - \frac{Q_\text{HII}}{t_\text{rec}}, 
    \label{eq:reionization}
\end{equation}
where $\dot{n}_\text{ion,H}$ is the emissivity of hydrogen-ionizing photons, $\langle n_\text{H}\rangle$ is the volume averaged hydrogen number density, and $t_\text{rec}$ is the average recombination time. The recombination time can be parameterized with an effective clumping factor $\mathcal{C}_R$, 
\begin{equation}
     1/t_\text{rec}=\chi_\text{e}\alpha_\text{A}(T)\langle n_\text{H}\rangle(1+z)^3\mathcal{C}_\text{R},
\end{equation} 
where $\chi_\text{e}=1.08$ is the number of electrons available for recombinations per hydrogen atom, and $\alpha_\text{A}(T)$ is the Case-A recombination rate. Given the hydrogen number density, gas temperature in ionized regions, emissivity, and the volume-filling factor of ionized hydrogen in our simulations as input, we can solve Equation ~\ref{eq:reionization} to infer the effective clumping factor. This clumping factor is shown in panel~C of Figure~\ref{fig:Clumping}. We also show estimates from \citet{Chen2020} and \citet{Davies2024}. 

Our four reionization models show a similar trend in the clumping factor evolution. The value is relatively flat at redshifts, where the IGM is still predominately neutral. As reionization proceeds, the clumping factor slightly decreases and then increases towards the end of reionization. We infer a lower clumping factor towards the end of reionization if the midpoint of reionization is at a higher redshift, as evident by the blue, green, and yellow curves. For our Early and Oligarchic models, the inferred effective clumping factor agrees with \citet{Chen2020}. For the Fiducial and Extremely Early models, it is somewhat higher and lower, respectively. 

The discrepancy with the clumping factor evolution suggested by \citet{Davies2024} (black squares in 
Figure~\ref{fig:Clumping}) is somewhat puzzling as \citet{Davies2024} derive their clumping factor values from \lya\ forest data that all of our simulations match very well. \citet{Davies2024} assumed a power-law for the dependence of  the mean free path  on frequency ($\lambda_\nu \propto \nu^{\alpha_\lambda}$)  and used an expression of the clumping factor in terms of the specific angle-averaged mean intensity which they assumed to depend on frequency as $J_\nu \propto \nu^{-\alpha_b}$, to write  
\begin{equation}
\mathcal{C}_\text{R}=\left \langle \frac{4\pi}{h}\frac{J_{\nu_\text{HI}}}{\lambda_{\nu_\text{HI}}}\left [\frac{1-4^{-\alpha_b-\alpha_\lambda}}{\alpha_b+\alpha_\lambda} \right ] \right \rangle \times \frac{1}{\chi_e\langle n_\text{H}x_\text{HII}\rangle^2 \langle\alpha_\text{{HII}}(T)\rangle},
\label{eq:Davies_CR}
\end{equation}
where $J_{\nu_\text{HI}}$ and $\lambda_{\nu_\text{HI}}$ are the values of $J_\nu$ and $\lambda_\nu$, respectively, at 912~\AA.\footnote{There is a typo in Equation~(16) of \citet{Davies2024} resulting in a dimensional inconsistency.  This has not affected their results, however. Their Equation ~(15) is correct.} \citet{Davies2024} then used measurements of the mean free path and the photoionization rate $\Gamma$ to estimate the effective clumping factor, where they related 
$\Gamma$  and $J_\nu$ as,
\begin{multline}
    \Gamma=4\pi\int_{\nu_\mathrm{HI}}^{4\nu_\mathrm{HI}} \frac{\mathrm{d}\nu}{h\nu} J_{\nu_\text{HI}}\left(\frac{\nu}{\nu_\text{HI}}\right)^{-\alpha_b}\\
=    4\pi \frac { J_{\nu_\text{HI}}}{h} \sigma_{\text{HI}}\left [\frac{1-4^{-\alpha_b-3}}{\alpha_b+3} \right ],  
\end{multline}
and they assume $\sigma_{\nu}=\sigma_\text{HI} (\nu/\nu_\text{HI})^{-3}$ where $\sigma_{\text{HI}}$  is the value of $\sigma_{\nu}$ at 912~\AA. 

Equation~\ref{eq:Davies_CR} helps us understand the discrepancy between the clumping factor values reported by \citet{Davies2024} with those in our simulations. There are two main differences. First, while evaluating Equation~\ref{eq:Davies_CR}, \citet{Davies2024} use the case~B recombination coefficient. In contrast, we use the case A recombination coefficient, which is about a factor 1.6 larger at the relevant temperatures and should be the correct choice in highly ionized regions at the tail-end of reionization  \citep{Madau2017}. Second, we note here that the scaling of the mean free path with photon frequency characterized by \citet{Davies2024} with $\alpha_{\lambda}$ is poorly constrained. This scaling should be sensitive to the contribution and the physical properties of Lyman-limit systems and the more diffuse gas, which are difficult to model and are currently poorly understood. Note further that the physical properties of the absorbers responsible for the Lyman-continuum opacity change rapidly at the tail end of reionization when the mean free path increases rapidly and that modelling this correctly will require high-resolution fully-coupled radiative transfer simulations with large numbers of frequency bins \citep{Madau2017, Feron2024}. Since our simulations have a small number of frequency bins and only marginally resolve Lyman-limit systems situated in galactic haloes, we can not properly account for the spectral hardening of the ionizing UV background.
Furthermore, as the mean free path is still limited by the remaining neutral islands \citep{Feron2024}, $\alpha_{\lambda}=3$ should be a more appropriate choice than $\alpha_{\lambda}=1$, used by \citet{Davies2024}. This further reduces $C_\text{R}$ by an additional factor $\sim 1.88$. The open symbols in Figure~\ref{fig:Clumping} show the effective clumping factor of \citet{Davies2024} corrected downward by a factor $\sim 3$ for these choices. The reduced clumping factors agree well with that inferred from our simulations in the redshift range where they overlap, as well as earlier estimates of the clumping factor from \lya\ forest data at $z=6$ \citep{Bolton2007}.

\section{Discussion and Conclusions} \label{sec:conclusion}

Matching the observed UV luminosity function at $5<z<12.5$ with the ionizing sources in our \atmf\ simulations and assuming the recent measurements of $\xi_\text{ion}$ by \citet{Simmonds2024b}, we have inferred the escape fractions of ionizing photons. We have also discussed the effective clumping factor inferred from our simulations that match \lya\ forest data 
at the tail-end of reionization. Our conclusions are as follows.

\begin{itemize}
    \item For our oligarchic source model, the inferred escape fraction rises from 10\% at $z=6$ to uncomfortably large values $>50$\% at $z>10$, disfavouring the oligarchic source model at very high redshift. With our fiducial source model, the inferred escape fractions rise from 5--6\% at $z=6$ to  15--30\% at $z=10$. The earlier reionization histories then require further rising escape fractions towards higher redshift, where the number density of dark matter haloes with masses above the resolution limit of the simulations rapidly decreases. The rise is more rapid if we extrapolate the observed luminosity function to $M_{\rm UV} =-11$  as $L^{-1}$, perhaps suggesting that the observations have not yet reached the faint-end turnover at high redshift.
    \item At the tail end of reionization, the escape fractions are in reasonable agreement with those observed at lower redshift, and at high redshift, the earlier reionization histories agree with the lower end of estimates inferred using the scaling relations from \citet{Chisholm2022}.
    \item For our Early reionization history, the effective clumping factor characterizing the number of recombinations thereby agrees well with that of  \citet{Chen2020}. It is somewhat 
    higher (lower) for our Fiducial (Extremely Early) model. Assuming case A recombination and 
    a scaling of the mean free path with frequency appropriate for our simulations, it also agrees well with recent estimates of the effective clumping factor by \citet{Davies2024} using \lya\ forest 
    data in the redshift range of overlap.
\end{itemize}

Overall our simulations show remarkable consistency between the ionizing properties of reionization-epoch galaxies reported by the JWST and the ionizing photon budget inferred from JWST observations and \lya\ forest data.

\section*{Acknowledgements}
SA and MGH thank Nick Gnedin, Harley Katz, Piero Madau,  Roberto Maiolino and Brant Robertson for helpful discussions at the KITP program ``Cosmic Origins: The First Billion Years" that informed this work. MH also thanks Prakash Gaikwad for helpful comments. The work was performed partially using the Cambridge Service for Data Driven Discovery (CSD3), part of which is operated by the University of Cambridge Research Computing on behalf of the STFC DiRAC HPC Facility (\href{www.dirac.ac.uk}{www.dirac.ac.uk}). The project was also supported by a Swiss National Supercomputing Centre (CSCS) grant under project ID s1114. This research was partly supported by grant NSF PHY-2309135 to the Kavli Institute for Theoretical Physics (KITP). Support by ERC Advanced Grant 320596 `The Emergence of Structure During the Epoch of Reionization’ is gratefully acknowledged. MGH has been supported by STFC consolidated grants ST/N000927/1 and ST/S000623/1. GK gratefully acknowledges support from the Max Planck Society via a partner group grant. GK is also partly supported by the Department of Atomic Energy (Government of India) research project with Project Identification Number RTI 4002. The work has been performed as part of the DAE-STFC collaboration `Building Indo-UK collaborations towards the Square Kilometre Array' (STFC grant reference ST/Y004191/1). SA also thanks the Science and Technology Facilities Council for a PhD studentship (STFC grant reference ST/W507362/1) and the University of Cambridge for providing a UKRI International Fees Bursary.

\section*{Data availability}
All data and analysis code used in this work are available from the first author upon request.

\bibliographystyle{mnras}
\bibliography{refs} 

\bsp	
\label{lastpage}

\end{document}